# Logarithmic Correlations For Turbulent Pipe Flow Of Power Law Fluids

**Trinh, Khanh Tuoc**

**Institute of Food Nutrition and Human Health**

**Massey University, New Zealand**

*K.T.Trinh@massey.ac.nz*

---

**ABSTRACT**

This paper uses the estimates of phase-locked parameters at the onset of bursting presented in a companion paper to derive logarithmic correlations for turbulent friction factor losses in time-independent power law fluids.

Two different techniques for analysis were used. They gave logarithmic correlations with different coefficients. But both correlations predicted 264 data points of friction factors published in the literature with a standard error of 1.3%.

The derivations show how modern understanding of coherent structures embedded in turbulent flow fields can be used to derive engineering correlations for head losses.

# 1. Introduction

In a companion paper (Trinh, 2009), it has been argued that turbulence is a local instantaneous phenomenon, which cannot be adequately explained by measurements and analysis of time-averaged shear stresses and velocities only. A great confusion has resulted from starting the theoretical analysis with the time-averaged Navier-Stokes equations whereas the proposed theory starts with the unsteady state Navier-Stokes equations and then time-averages the solution.

The flow field was divided into a wall layer, where viscous effects are important and an outer region where they are not (Panton, 1990). The physical model for the wall layer shown in Figure 1 was based on the classic observations of (Kline et al., 1967).

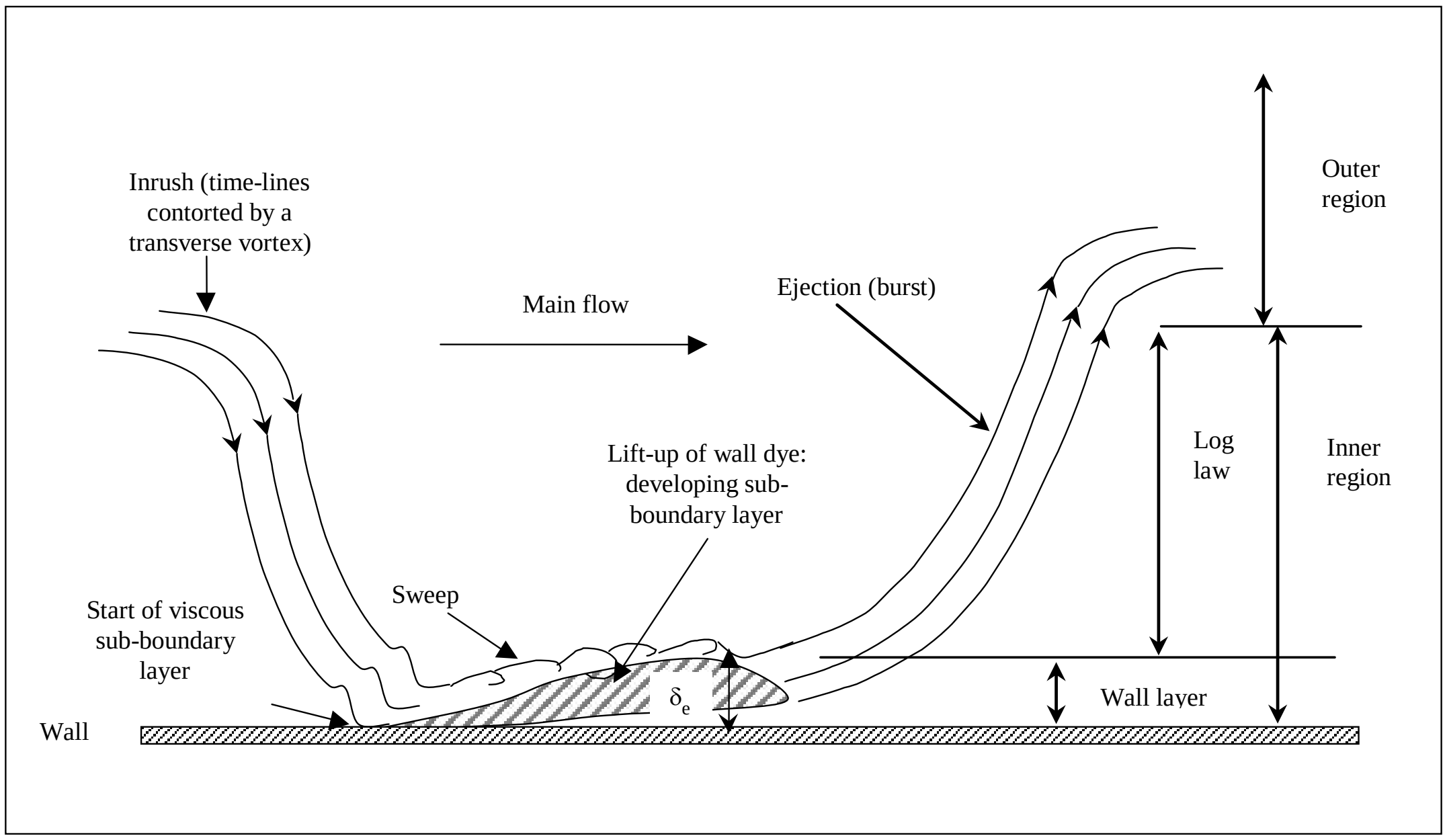

Figure 1. A cycle of the wall layer process. Drawn after the observations of Kline *et. al.* (1967).

Despite the dominating influence of viscous momentum, the flow field near the wall is not laminar in the steady-state sense, but highly active. Periodically, fast fluid rushes from the outer region towards the wall. This fluid is deflected into a vortical sweep along the wall. The travelling vortex induces a viscous sub-boundary layer underneath its path . The viscous sub-boundary layers are observed as streaks of low-speed fluid, which tend to lift, oscillate and eventually burst in a violent ejection from the wall towards the outer region. The low-speed-streak phase is much more persistent than the ejection phase and dominates the contribution to the time-averaged profile (Walker et al., 1989).

The edge of the wall layer was defined as the position of maximum penetration of wall retardation through viscous exchange of momentum between the wall and the fluid stream.

This corresponds to the thickness of the transient sub-boundary layer at the point of bursting, $\delta_e$, as indicated in Figure 1. The critical instantaneous wall shear stress at the point of bursting, $\tau_e$, was estimated from measurements of the time-averaged wall shear stress, $\tau_w$.

Plots of the critical instantaneous friction factor-Reynolds Number for time-independent power law and Newtonian fluids coincide completely. Prandtl's logarithmic law (Nikuradse, 1932)

$$f^{-(1/2)} = 4.00\log(\mathrm{Re}\ f^{1/2}) - 0.40 \qquad (1)$$

was rewritten in terms of the critical wall shear stress

$$f_e^{-(1/2)} = 5.66\log(\mathrm{Re}_e f_e^{1/2}) + 0.45 \qquad (2)$$

Equation (2) correlated 264 data points published in the literature with a range of Reynolds number from 4000 to 240,000 and flow behaviour index from 0.214 to 1.00 with a standard deviation of less than 3%. This result indicated that

1. The mechanism of turbulence generation near the wall is the same for time independent non-Newtonian and Newtonian fluids. The natural normalization parameters for turbulent flow are the local critical instantaneous (phase-locked) parameter at the onset of bursting. The transformation from time-averaged shear stress measurements to local instantaneous shear stresses was made through the application of a Lagrangian penetration theory for viscous momentum (Trinh, 2010, Trinh and Keey, 1992a).
2. The difference between friction factor plots for power law and Newtonian fluids reported previously in the literature may be attributed to the use of time-averaged wall shear-stress measurements, which introduce an additional integration coefficient. It does not prove the existence of separate mechanisms of turbulence generation for Newtonian and non-Newtonian fluids as has been argued, for example by (Dodge and Metzner, 1959, Tennekes, 1966).

However, engineers require time-averaged head losses more than local instantaneous friction factor plots, which are more helpful to explain fundamental theoretical concepts.

This paper shows how the new fundamental theory of turbulence in non-Newtonian time independent fluids can be applied to derive logarithmic correlations for time averaged friction factors. Power law correlations of the Blasius type are derived in a third paper in this series. Two different techniques will be applied to derive the logarithmic correlations. They give predictions of similar accuracies and can help elucidate misconceptions about the relationship between correlations for the friction factor and the velocity profile in time-independent non-Newtonian fluids.

This derivation is made for fluids obeying the power law, also called Ostwald De Waele, rheological model:

$$\tau = K\dot{\gamma}^n \qquad (3)$$

where: $\tau$ the shear stress,
$\dot{\gamma}$ the shear rate
K the consistency coefficient, and
n the flow behaviour index

## 2. Derivation of a logarithmic correlation using Karman's method[11]

The turbulent core in Newtonian fluids can be described by the well-known logarithmic law-of-the-wall (Nikuradse, 1932):

$$U^{+} = 2.5 \ln y^{+} + B \tag{4}$$

Bogue and Metzner (1963) have argued that this equation also describes the turbulent core in time-independent non-Newtonian turbulent flow. Their velocity data tend to support this argument but the accuracy of these measurements, which are difficult, was not high.

The non-dimensional distance $y^+$ is a kind of local Reynolds number:

$$y^{+} = \frac{y u_*}{\nu_w} = \frac{y u_*^{2/n-1} \rho^{1/n}}{K^{1/n}} \tag{5}$$

where $u_*$ is called the friction velocity

$$u_* = \sqrt{\frac{\tau_w}{\rho}} \tag{6}$$

and $\nu_w$ is the apparent non-Newtonian viscosity at the wall:

$$\nu_w = \frac{\tau_w}{\rho \dot{\gamma}_e} = \frac{K^{1/n} \tau_w^{(n-1)/n}}{\rho} \tag{7}$$

The non-dimensional local time-averaged velocity $U^+$ is a kind of local friction factor

$$U^{+} = \frac{u}{u_*} \tag{8}$$

The distance and local velocity may also be normalised with the critical instantaneous wall shear stress at the point of bursting (Trinh, 2009)

$$\tau_e = \frac{\tau_w}{n+1} \tag{9}$$

and

$$u_{e*} = \sqrt{\tau_e / \rho} = \sqrt{\frac{\tau_w}{\rho(n+1)}} = \frac{u_*}{\sqrt{n+1}} \tag{10}$$

The constant B in equation (4) is defined if any point on the velocity profile is known. One of the most convenient points to estimate is the edge of the wall layer described by the point ($U_e$,$\delta_e$). The wall parameters can be normalised in two ways, with the critical friction velocity:

$$U_e^+ = \frac{U_e}{u_{e*}} \tag{11}$$

$$\delta_e^+ = \frac{\delta_e u_{e*}}{\nu_e} = \frac{\delta_e u_{e*}^{2/n-1} \rho^{1/n}}{K^{1/n}} \tag{12}$$

where

$$\nu_e = \frac{\tau_e}{\rho \dot{\gamma}_e} = \frac{K^{1/n} \tau_e^{(n-1)/n}}{\rho} = \frac{K^{1/n} (n+1)^{(n-1)/n} \tau_w^{(n-1)/n}}{\rho} \tag{13}$$

or with the time-averaged shear velocity:

$$U_\nu^+ = \frac{U_e}{\sqrt{\tau_w / \rho}} = \frac{U_e^+}{\sqrt{n+1}} \tag{14}$$

and

$$\delta_\nu^+ = \frac{\delta_e u_*^{2/n-1} \rho^{1/n}}{K^{1/n}} = \delta_e^+ (n+1)^{2-n/2n} \tag{15}$$

The relationship between the wall layer approach velocity and thickness was found to be independent from the flow behaviour index when normalised with the critical shear velocity

$$\delta_e^+ = 2.08 U_e^+ \tag{16}$$

Hence it applies to both Newtonian and power law fluids. However, the relationship normalised with the time averaged shear velocity does not

$$\delta_\nu^+ = 2.08 (n+1)^{1/n} U_\nu^+ \tag{17}$$

Equation (4) passes through the point ($U_\nu^+$, $\delta_\nu^+$) defined by equations (14) and (15). Thus the coefficient B is obtained from equation (4) as:

$$B = U_\nu^+ - 2.5 \ln \delta_\nu^+ \tag{18}$$

For a Newtonian fluid, n=1, Nikuradse reported that B=5.5. Substituting into equations (17) and (18) give $U_\nu^+$ = 16.2 and $\delta_\nu^+$= 67. Substituting these values into equations (14) and (15) and putting n=1 gives $U_e^+$ = 22.6 and $\delta_e^+$= 47.4.

Substitution of these values of $Ue^+$ and $\delta_e^+$ back into equations (14) and (15) gives general parameters for the wall layer in power law fluids and the coefficient B becomes:

$$B = 16.2 (n+1)^{1/2} - 2.5 \ln 67 (n+1)^{(2-n)/2n} \tag{19}$$

Substituting equation (19) into (4) gives the law of the wall for a power law Ostwald De Waele fluid in turbulent pipe flow as

$$U^{+} = 2.5\ln y^{+} + 16\,(n+1)^{-1/2} - 9.64 - \frac{2.5(2-n)}{2n}\ln(n+1) \quad (20)$$

The friction factor is defined as

$$f = \frac{2\tau_w}{\rho V^2} \quad (21)$$

Metzner and Reed (1955) defined a generalised Reynolds number, which is widely used:

$$\mathrm{Re}_g = \frac{D^n V^{2-n}\rho}{K\,8^{n-1}\left(\frac{3n+1}{4n}\right)^n} \quad (22)$$

The relation between maximum velocity $U_m^+$, at the pipe centre $R^+$, and the average discharge velocity $V^+$ in Newtonian fluids

$$U_m^+ = V^+ + Q(n) \quad (23)$$

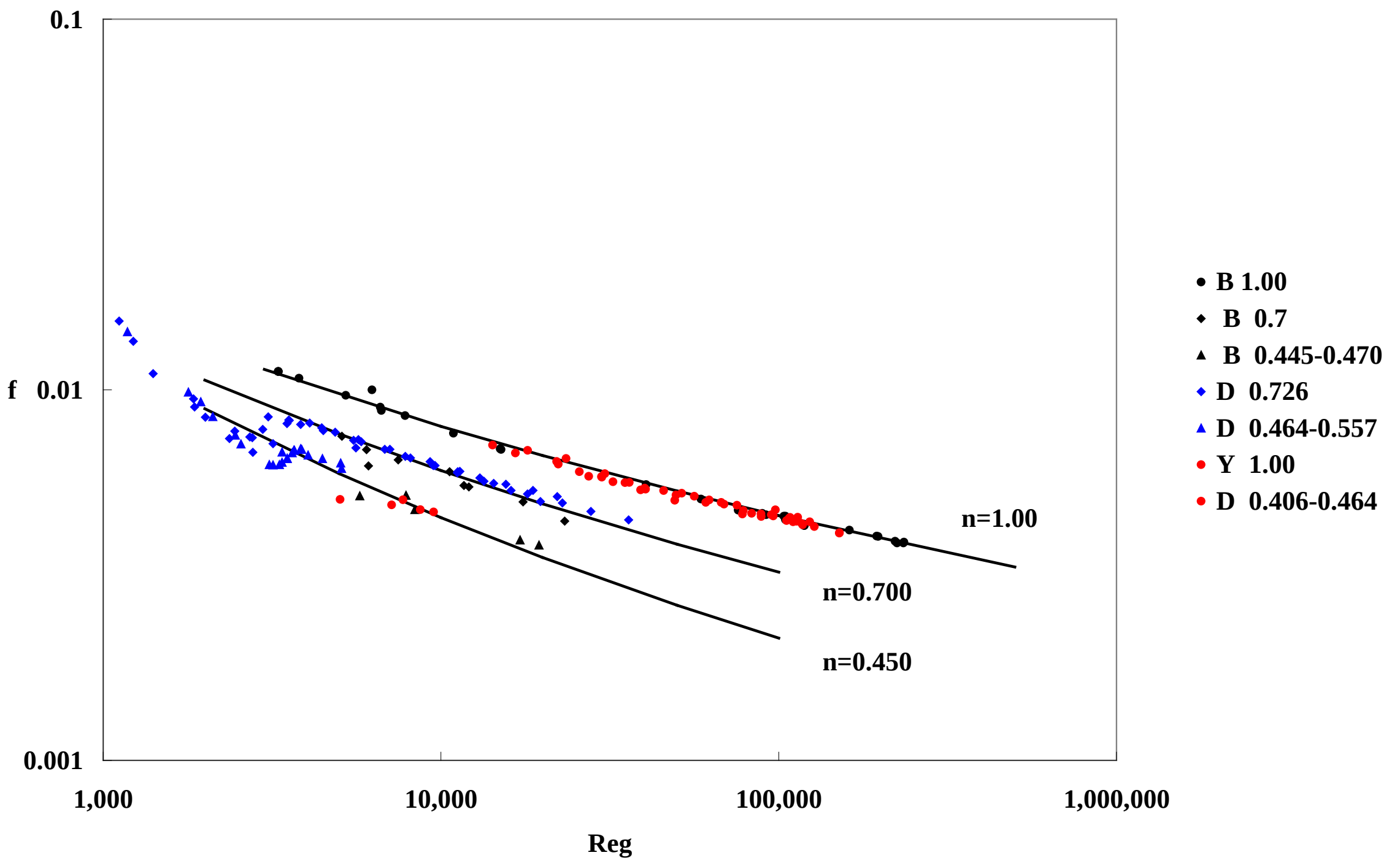

Figure 2. Friction factor plot for three typical values of flow behaviour index. Data source B (Bogue, 1962), D (Dodge, 1959), Y (Yoo, 1974). Number represent ranges of $n$ values

The parameter $Q(n)$ for Newtonian fluids is given by Nikuradse as $Q(1) = 4.03$ and can be calculated for each value of $n$ by numerical integration using from equation (20). But the Nikuradse value may be used for power law fluids with little error because the wall layer is thin compared with the outer region where equation (4) applies to both Newtonian and power law fluids. Setting $y^+ = R^+$ in equation (20) and combining with equations (21), (22) and (23) gives a relation for the friction factor:

$$\frac{1}{\sqrt{f}} = \frac{4.07}{n}\log(Re_g f^{1-n/2}) + 11.31\left(\frac{2}{n+1}\right)^{1/2} - 5.99 - \frac{6.13}{n} - \frac{4.07}{n}\log\left[\left(\frac{n+1}{2}\right)^{\frac{2-n}{2}} \Big/ \left(\frac{3n+1}{4n}\right)^{n}\right] \quad (24)$$

Equation (24) predicts 264 friction factor data points measured by Dodge, Bogue and Yoo with a standard error of 1.3%. The data covers a range of flow behaviour index between 0.214 and 1 and Metzner-Reed Reynolds number between 4000 and 220,000. The effect of the flow behaviour index is illustrated in Figure 2

The ratio $R_f$ of predicted-to-measured friction factors should ideally be unity. Figure 3 shows that the predictions hold well for the full range of Reynolds numbers.

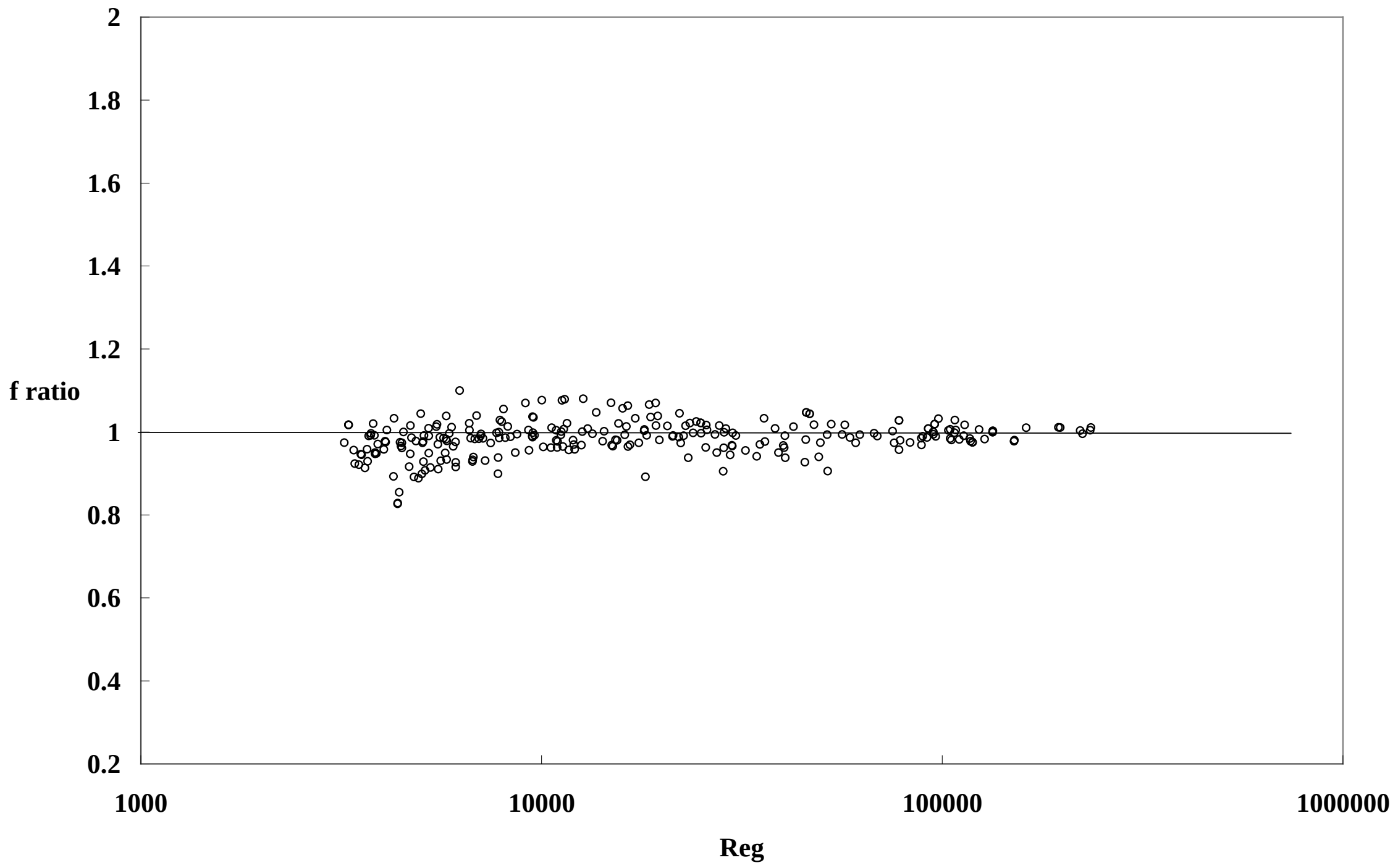

Figure 3. Ratio of predicted to measured friction factors against Reynolds numbers for all values of n.

## 3. Derivation from the critical friction factor master curve

The critical instantaneous friction factor and Reynolds number are defined as[1]

$$f_e = \frac{2\tau_e}{\rho V^2} \qquad \mathrm{Re}_e = \frac{DV}{\nu_e} \tag{25}$$

They are related to the time-averaged friction factor and Metzner-reed Reynolds number by:

$$f = f_e(n+1) \tag{26}$$

$$\mathrm{Re}_e = \left(\mathrm{Re}_g f^{1-n}\right)^{1/n} 2^{59n-1)/n} \left(\frac{(3n+1)}{4n}\right)\left(\frac{(n+1)}{2}\right)^{1-1/n} \tag{27}$$

Substituting equations (25), (26) and (27) into equation (2) gives

$$\frac{1}{\sqrt{f}} = \frac{5.67}{n\sqrt{n+1}}\log\left(\mathrm{Re}_g f^{1-n/2}\right) + \frac{5.66(n-1)}{n\sqrt{n+1}}\log\left(2^4(n+1)\right) - \frac{5.66}{2\sqrt{n+1}}\log(n+1) + \frac{5.66}{\sqrt{n+1}}\log\left(\frac{3n+1}{4n}\right) + \frac{0.45}{\sqrt{n+1}} \tag{28}$$

Equation (28) also correlates the 264 data points reported previously very well as shown in Figure 4. The standard error is 1.3%.

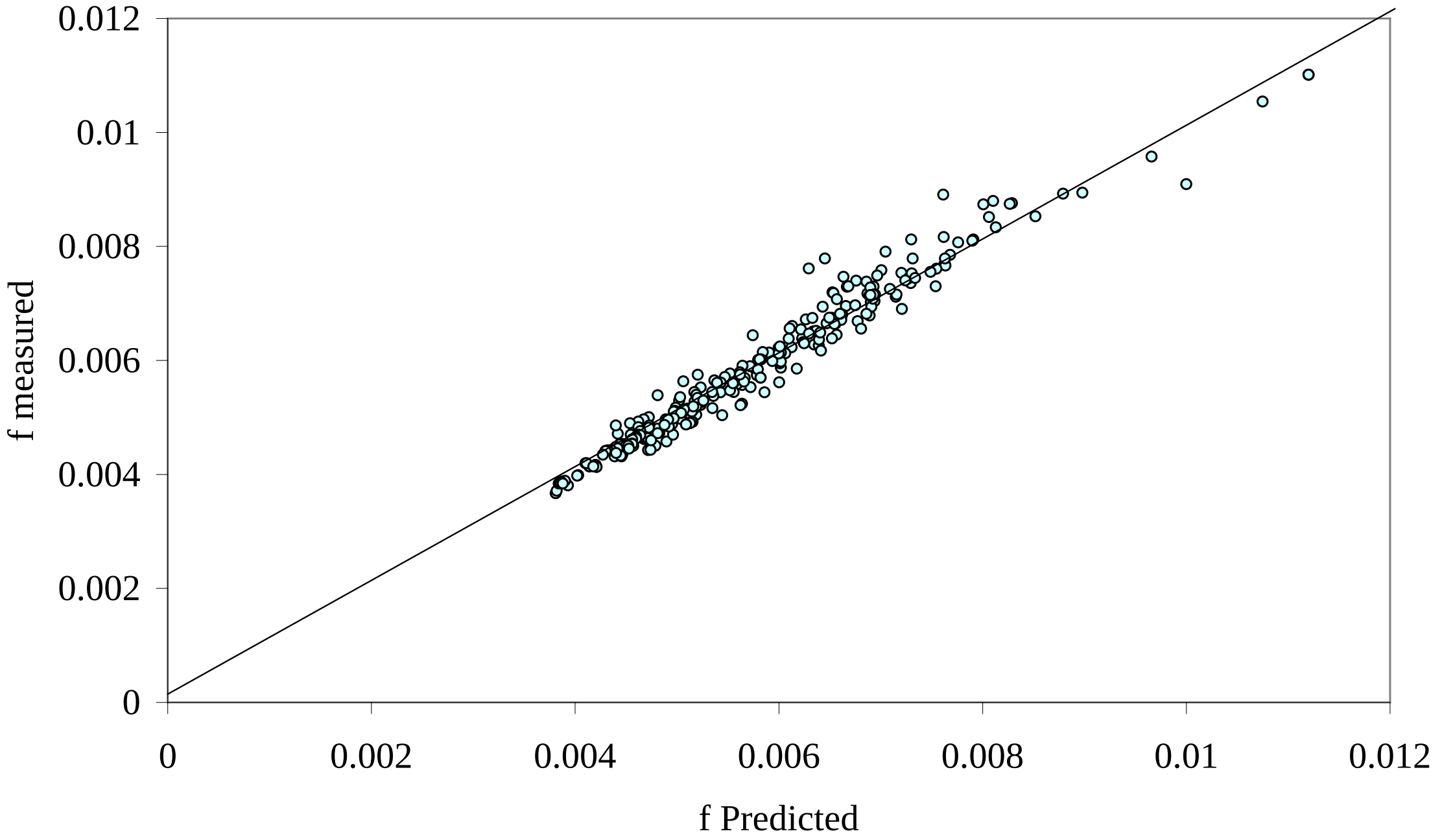

Figure 4. Comparison of measured friction factors and predictions from equation (28).

## 4. Discussion

Dodge and Metzner (1959) derived a general form of the logarithmic friction factor-Reynolds number correlation by dimensional arguments

$$\frac{1}{\sqrt{f}} = A_n \mathrm{Log}\left(\mathrm{Re}_g f^{1-n/2}\right) + B_n \tag{29}$$

The coefficients $A_n$ and $B_n$ were obtained empirically from experimental measurements of friction factor to give

$$\frac{1}{\sqrt{f}} = \frac{4}{n^{0.75}} \text{Log}(\text{Re}_g\, f^{1-n/2}) - \frac{0.4}{n^{1.2}} \tag{30}$$

Tennekes (1966) proposed a theoretical explanation for the value of $A_n$ by considerations of eddy scales in the inertial subrange of the turbulent spectrum. Dodge and Metzner also derived a correlation for the velocity profile based on equation (30)

$$u^+ = \frac{5.66}{n^{.75}} \log y^+ - \frac{0.566}{n^{1.2}} + \frac{3.475}{n^{.75}}\left[1.960 + 0.815n - 1.620n\log\left(3+\frac{1}{n}\right)\right] \tag{31}$$

However, equation (31) does not match the velocity profiles measured by Bogue (op.cit.). Bogue and Metzner (1963) showed that the slope of the $U^+$- $\ln(y^+)$ curve has the same value of 2.5 for both Newtonian and power law fluids. Despite this inconsistency between works from the same school, equation (30) has been considered the standard correlation for non-Newtonian turbulent pipe flow in the last fifty years and has been cited 329 times.

The present analysis helps to explain this apparent paradox. Two correlations, equations (24) and (28), were derived from the same conceptual model of wall turbulence. For the same velocity profile, the coefficients $A_n$ in these equations were different functions of the flow behaviour index n but both equations predicted experimental data with equal success. Thus the empirical correlation of Dodge and Metzner, equation (30), does not require that the slope of the logarithmic law of the wall be different for Newtonian and power law fluids. Nor does it imply difference in the mechanism of wall turbulence as argued by Tennekes. Indeed, the numerical predictions from logarithmic correlations for friction factors are relatively insensitive to small differences in values for the coefficient $A_n$, even in Newtonian fluids, e.g. Karman (1931), Nikuradse (1932)and Laufer (1954). The derivation of a correlation for the friction factor like equation (24) from a semi-empirical model for the velocity profile like equation (20) is more reliable than the reverse exercise.

The understanding of turbulence has improved remarkably in the last fifty years. Since Kline *et. al.* published their ground-breaking visualisation of the structure of the wall layer and the evolution of the low-speed-streaks through an inrush-sweep-burst/ejection cycle, there has been an explosion of research on coherent structures embedded in turbulent flow fields. The successful development of direct numerical simulations (DNS) two decades ago has allowed us to map in extraordinary details all the transient structures found in all regions of turbulent flow fields.

The striking observation for people like me, who only came out late from a confined and isolated scientific environment, is that none of this modern understanding has permeated into predictive correlations of friction losses used in engineering design. These are still based on concepts proposed in the early 1930's. The correlation of Dodge and Metzner remains the most trusted for predictions of turbulent losses in power law fluids.

The analysis presented in this series of paper attempts to bridge that gap. The remarkable observations of Kline *et. al.* has clearly stressed the need to analyse turbulence production with the original unsteady state Navier-Stokes equations then average the solution to give engineering correlations. To begin the analysis with the time-averaged Navier-Stokes equations, the famous Reynolds equations, adds more confusion to the picture.

However, the solution of the Navier-Stokes equations for the entire turbulent flow field is a daunting task. The present analysis has adopted a zonal approach. By focusing on a key structure, the transient viscous sub-layer in the low-speed-streak underneath the longitudinal vortex near the wall, the Navier-Stokes can be greatly simplified. This solution can then be time-averaged and matched with well-accepted solutions for the outer region, like the logarithmic law of the wall, to give correlations for the entire field.

Einstein and Li (1956)and Meek and Baer (1970) were among the first to have used this technique for Newtonian fluids. They modelled the wall layer after Stokes' first problem of a flat plate suddenly set in motion in a fluid

$$\rho \frac{\partial u}{\partial t} = - \frac{\partial \tau}{\partial y} \tag{32}$$

There are two issues with this approach:

1. Stokes' solution is Eulerian. It should include two further convection terms $u(\partial u/\partial x)$ and $v(\partial u/\partial y)$ but these were omitted on the grounds that they are not important in an impulsively started motion. The validity of this simplification to the wall layer structure has never been tested.

   The convection terms drop out naturally in a Lagrangian model (Trinh and Keey, 1992a, Trinh and Keey, 1992b, Trinh, 2010) based on a new class of partial derivatives following the motion of the diffusing species.

2. Inrushes, and therefore low-speed-streaks, occur at random positions on the wall surface. Thus a probe positioned at a fixed position near the wall will see many different low-speed-streaks over a period of time. Thus time averaging the solution for a single viscous sub-layer may not necessarily give a good estimate of the time average velocity or wall shear stress measured by the fixed probe. The two coincide only if one assumes that the low-speed-streaks at all stages of development pass the fixed probe with equal probability (Trinh, 1992).

The present analysis has identified a particularly important phase-locked parameter, the critical friction factor $f_e$ at the onset of bursting or ejection. This factor turned out to be the same for both Newtonian and time-independent power law fluids. The proof has to be shown separately for each other rheological model. Solutions for the most common will be presented in other publications.

## 5. Conclusion

The phase-locked parameters at the onset of bursting have been successfully used to give logarithmic correlations for turbulent friction factor losses in time-independent power law fluids. The predictions rival in accuracy those made with the Dodge-Metzner correlation.

In addition, the physical visualisation provides an intuitively satisfying explanation for the mechanism of turbulence production in power law and Newtonian fluids. A definitive proof can only be obtained by repeating the work of Kline et al for power fluids and measuring the actual phase locked parameters.

While this is a big task yet to be performed, the methodology developed here can provide a sound framework for modelling of non-Newtonian turbulent flow in other geometries and for other rheological models.

## 6. Nomenclature

| | |
|---|---|
| $A_n$, $B_n$ | Empirical coefficients defined by Dodge in equation (29) |
| B | Coefficient defined in equation (4) |
| D | Pipe diameter |
| f | Time-averaged friction factor defined by equation (21) |
| $f_e$ | Critical (local instantaneous) friction factor defined by equation (25) |
| K | Consistency coefficient in power law model |
| n | Flow behaviour index in power law model |
| Re | Reynolds number, DV/ν |
| $Re_e$ | Critical instantaneous Reynolds number, $DV/\upsilon_e$, equation (25) |
| $Re_g$ | Metzner-Reed generalised Reynolds number, equation (22) |
| t | Time |
| U | Local time-averaged velocity |
| u | Local instantaneous velocity smoothed with respect to fluctuations imposed by a travelling vortex |
| $u_*$ | Time-averaged friction velocity |
| $u_{e*}$ | Critical (instantaneous) friction velocity |
| $U_e$ | Approach velocity to transient viscous sub-boundary layer (velocity at wall layer edge at the moment of ejection) |
| $U_m$ | Maximum velocity (at pipe axis) |
| $U_v^+$ | Approach velocity normalised with time-averaged friction velocity, $U_e/u_*$ |
| $U_e^+$ | Approach velocity normalised with the critical shear velocity, $U_e/u_{e*}$ |
| V | Mixing cup average discharge velocity |
| y | Normal distance from the wall |
| $y^+$ | Normalised distance defined by equation (5) |
| $\delta_e$ | Instantaneous thickness of (Stokes) transient sub-boundary layer |
| $\delta_e^+$ | Wall layer normalised with the critical shear velocity |
| $\delta_v^+$ | Wall layer thickness normalised with the time-averaged shear velocity |
| $\dot{\gamma}$ | Shear rate |
| $\nu_e$ | Apparent critical kinematic viscosity defined by equation (13) |

ρ Density
τ Shear stress

Subscripts

ν Viscous, or at the edge of the wall layer
e At the onset of ejection or at the end of the low-speed streak phase, see also T
i instantaneous
T At the end of the low-speed-streak phase
w Wall

Superscript

\+ Normalised with wall shear stress